\documentclass[
reprint,
amsmath,
amssymb,
aps,
longbibliography,
floatfix
]{revtex4-2}

\usepackage{graphicx}
\usepackage{xcolor}
\usepackage[english]{babel}
\usepackage[T1]{fontenc}
\usepackage[utf8]{inputenc}
\usepackage{dcolumn}
\usepackage{bm}
\usepackage{hyperref}
\usepackage[capitalise]{cleveref}
\usepackage[mathlines]{lineno}
\usepackage{multirow}
\usepackage{braket}

\usepackage{MnSymbol}
\usepackage{physics}

\begin{document}

\preprint{APS/123-QED}

\title{
Multiparticle entanglement in a time-dependent electron--ion--photon system
}

\author{Axel Stenquist
}
\author{Jan Marcus Dahlström$^{{\text{\dagger}}}$
}

\affiliation{Department of Physics, Lund University, 22100 Lund, Sweden.}

\begin{abstract}
Photoionization provides a natural setting for entanglement transfer in the time domain: an outgoing electron can become entangled with the residual ion, while subsequent spontaneous emission may transfer these correlations to an emitted photon. We quantify the entanglement dynamics of the composite electron--ion--photon system using the von Neumann entropy of the one-versus-rest bipartitions, showing that the initially entangled electron--ion system evolves through a genuinely multiparticle-entangled state before approaching an entangled electron--photon state. Logarithmic negativity further resolves the pairwise entanglement, revealing in particular transient ion--photon entanglement. Further, its enhanced sensitivity to weakly populated Schmidt modes additionally exposes high-dimensional entanglement associated with the photon-energy continuum.
\end{abstract}

\maketitle

\begin{table}[b!]
\begin{flushleft}
  $^\text{\dagger}$ marcus.dahlstrom@fysik.lu.se
\end{flushleft}
\end{table}

\begin{figure}
    \centering
    \includegraphics[width=1\linewidth]{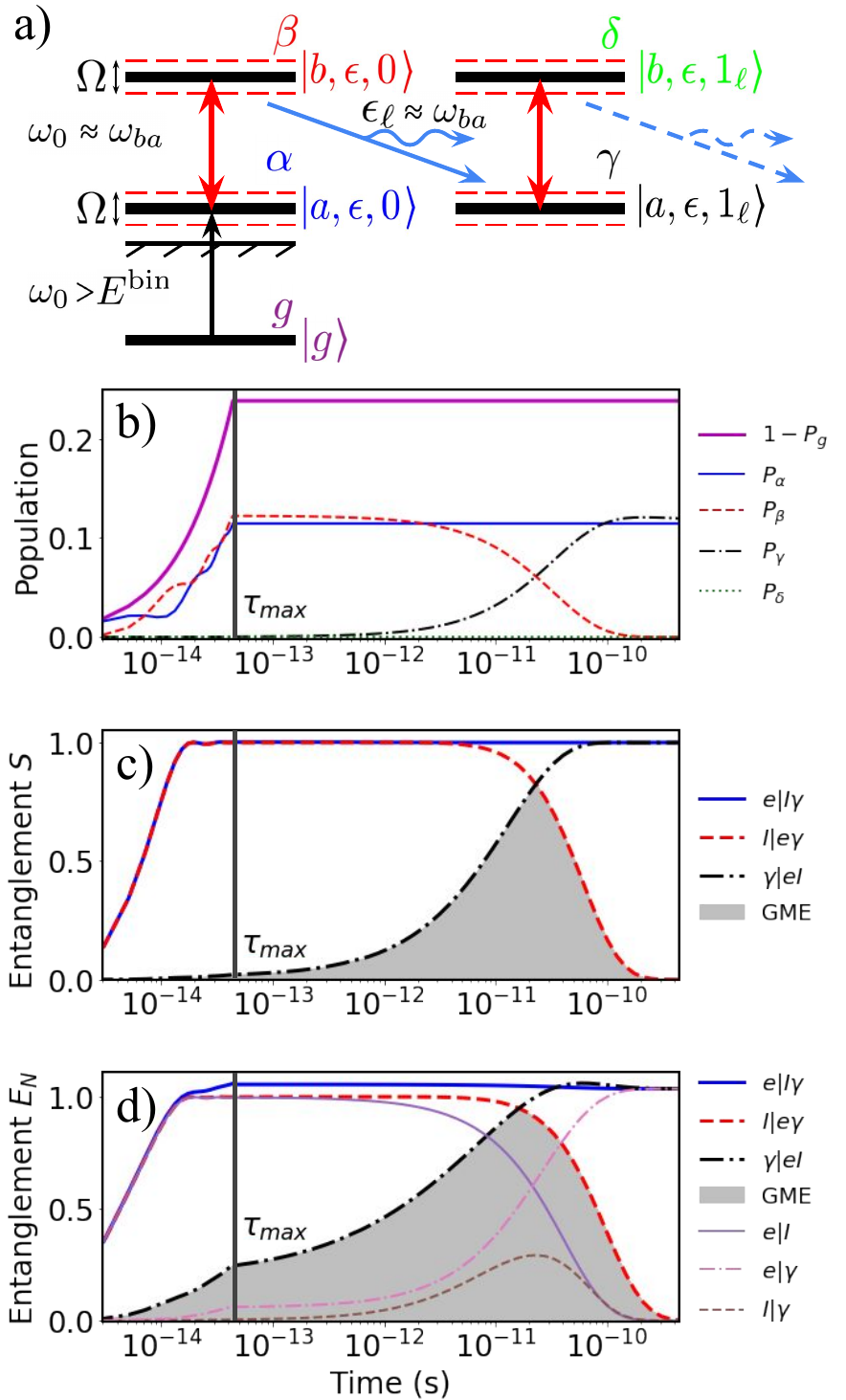}
    \caption{
    \textbf{Entanglement transfer in the conditioned electron--ion--photon system.}
    (a) Level scheme illustrating the wave function in Eq.~(\ref{Eq WF}). The driving field ionizes the atom and couples ionic states $a$ and $b$, while spontaneous emission transfers population from $\ket{b,\epsilon,0}$ to $\ket{a,\epsilon,1_\ell}$. Subsequently, the driving field strongly couples the states $\ket{a,\epsilon,1_\ell}\leftrightarrow\ket{b,\epsilon,1_\ell}$. Due to the ultrashort laser interaction higher-order spontaneous-emission processes are neglected.
    (b) Population dynamics of the ionized wave-function components and the total ionized population $1-|g(t)|^2$.
    (c) Entanglement of the three one-versus-rest bipartitions quantified by the von Neumann entropy. The gray region shows $\mathcal{E}_{\mathrm{GME}}$ from Eq.~(\ref{eq:MPE}).
    (d) Corresponding logarithmic negativities, together with the pairwise electron--ion, electron--photon, and ion--photon negativities. 
    }
    \label{fig:1}
\end{figure}

Entanglement may be redistributed between the constituents of a composite quantum system without being destroyed, forming the basis for entanglement transfer and quantum-state transfer protocols \cite{banchi_optimal_2010,giddings_quantum_2018}. Photoionization provides a natural setting for studying this process in the time domain: the outgoing electron can become entangled with residual ionic degrees of freedom in atoms \cite{laurell_continuous-variable_2022,mehmood_ionic_2023,nandi_generation_2024,stenquist_harnessing_2025}, as well as with electronic and nuclear degrees of freedom in molecules \cite{vrakking_control_2021,vrakking_ion-photoelectron_2022,ruberti_quantum_2022,makos_entanglement_2025,koll_entanglement_2026}; see also Ref.~\cite{tichy_essential_2011} for a broader perspective. Radiative decay may subsequently introduce a third, photonic, subsystem \cite{mollow_pure-state_1975}. In our previous work, we showed that fluorescence can transfer entanglement from an electron--ion pair to an electron--photon pair and identified an intermediate regime of multiparticle correlations \cite{stenquist_entanglement_2025}. However, the genuine multiparticle-entangled regime was not explicitly quantified. 
While photoionization-induced decoherence has been studied in the ionic \cite{goulielmakis_real-time_2010,pabst_decoherence_2011} and photoelectron subsystems \cite{busto_probing_2022,laurell_continuous-variable_2022,laurell_measuring_2025}, the three-particle nature of our scheme has unique properties. 

In this letter, two advances in the interpretation of composite electron-ion-photon states are presented. First, we explicitly compute the genuine multiparticle entanglement for pure states using the von Neumann entropy \cite{haroche_exploring_2006} and compare it to the corresponding logarithmic negativity \cite{vidal_computable_2002}. Second, we extend the analysis to mixed states, 
which allows us to regard any of the three particles as an environment that disrupts the coherent dynamics of the other two particles and alters the dynamics of their quantum entanglement. More explicitly, this allows us to distinguish entanglement of one particle with the remaining two from the pairwise entanglement retained when the third particle is unresolved.

The system and the relevant amplitudes are illustrated in Fig.~\ref{fig:1}(a). A short intense pulse ionizes the atomic ground state $\ket{g}$ and subsequently strongly couples two ionic states: $\ket{a} \leftrightarrow \ket{b}$ \cite{grobe_observation_1993,walker_observation_1995,zhang_photoemission_2014,yu_core-resonant_2018,nandi_generation_2024, stenquist_harnessing_2025}. On a timescale longer than this ultrafast laser interaction, $\tau_\text{sp} > \tau_\text{max}$, the excited ionic state decays radiatively, producing a photon in mode $\ell$, with energy $\epsilon_\ell$. Due to these different timescales, the probability for spontaneous emission of two photons is small. Thus, we restrict the radiation field to the vacuum and one-photon sectors, yielding the truncated wave function
\begin{equation}\label{Eq WF}
    \begin{split}
        \ket{\Psi(t)} \approx {}&
        g(t)\ket{g}
        +\int d\epsilon\,
        \Big\{
        \alpha(t,\epsilon)\ket{a,\epsilon,0}
        +\beta(t,\epsilon)\ket{b,\epsilon,0}
        \\
        &+
        \int d\epsilon_\ell
        \left[
        \gamma(t,\epsilon,\epsilon_\ell)\ket{a,\epsilon,1_\ell}
        +
        \delta(t,\epsilon,\epsilon_\ell)\ket{b,\epsilon,1_\ell}
        \right]
        \Big\}.
    \end{split}
\end{equation}
Here $\epsilon$ denotes the photoelectron energy and $\ket{1_\ell}$ a single photon in mode $\ell$. The calculation of the amplitudes $g$, $\alpha$, $\beta$, $\gamma$, and $\delta$ follows Ref.~\cite{stenquist_entanglement_2025} and is not repeated here. The ion--photon dynamics are described using the adiabatic-elimination approach applied by Mollow \cite{mollow_pure-state_1975}. This yields an effective non-Hermitian treatment of spontaneous emission, which accounts for irreversible decay to increasing photon numbers, as detailed in Ref.~\cite{stenquist_entanglement_2025}.

Figure~\ref{fig:1}(b) shows the corresponding population dynamics. Ionization occurs during the ultrashort Gaussian driving pulse, the pulse duration of which is scanned up to the maximal value $\tau_\text{max}=44$ fs, denoted by the vertical line. At the end of the pulse the $\alpha$ and $\beta$ branches have approximately equal populations. On the much longer radiative time scale, the excited-ion contribution $\beta$ decays while the one-photon amplitude $\gamma$ grows correspondingly. The $\alpha$ population remains unchanged and $\delta$ remains small. Spontaneous emission thus irreversibly converts the branch $\ket{b,\epsilon,0}$ into $\ket{a,\epsilon,1_\ell}$. 

We condition Eq.~(\ref{Eq WF}) on ionization by projecting out the neutral contribution $g(t)\ket{g}$ and renormalizing the remaining wave function. No further conditioning is performed: both ionic states, the complete electron spectrum, and both the vacuum and one-photon sectors are retained. The calculated entanglement consequently characterizes the complete ionized ensemble. This distinction is important because conditioning is a non-unitary operation and may alter the entanglement of the selected state. Additional postselection on, for example, the ionic state or photon number would lead to different quantum states, which may correspond to different degrees of entanglement.


Within the complete Hilbert space, the post-pulse evolution factorizes as $U\simeq U_e\otimes U_{I\gamma}$, which is local with respect to the bipartition $e|I\gamma$. The electron entanglement with the combined ion--photon system must therefore remain unchanged. In contrast, the ion--photon evolution cannot be factorized as $U_{I\gamma}\neq U_I\otimes U_\gamma$. Spontaneous emission couples the two subsystems through $\ket{b,0}\rightarrow\ket{a,1_\ell}$. In the effective non-Hermitian description this is represented by the term $\ket{a,1_\ell}\!\bra{b,0}$, while the reverse coupling $\ket{b,0}\!\bra{a,1_\ell}$ is eliminated. The ion--photon dynamics can therefore redistribute entanglement between the two subsystems.

For a pure bipartite state with density matrix $\rho_{AB}$, where $A$ and $B$ denote the subsystems, entanglement can be quantified by the von Neumann entropy 
\begin{equation}
S_{A|B}\equiv -\mathrm{Tr}\!\left(\rho_A\log_2\rho_A\right),
\label{eq:measures}
\end{equation}
where $\rho_A$ is the reduced density matrix of $A$ (or $B$). 
Because the conditioned electron--ion--photon state is pure, $S_{e|I\gamma}$, $S_{I|e\gamma}$, and $S_{\gamma|eI}$ are entanglement measures for the corresponding bipartitions.
This is no longer true for the pairwise states, attained by tracing out the third particle, for example $\rho_{eI}=\mathrm{Tr}_\gamma\rho_{eI\gamma}$. 
Because these reduced pairwise states are generally mixed, their von Neumann entropy quantifies mixedness rather than entanglement alone and is therefore not an entanglement measure. Their entanglement may instead be quantified using logarithmic negativity,  
\begin{equation}
    E_N(A|B)=\log_2\left\|\rho_{AB}^{T_A}\right\|_1,
\end{equation}
where $T_A$ is the partial transpose of subsystem $A$. The logarithmic negativity therefore allows all three pairwise combinations $e|I$, $e|\gamma$, and $I|\gamma$ to be investigated. 
We additionally apply the multiparticle entanglement indicator
\begin{equation}
\mathcal{E}_{\mathrm{GME}}
=
\min\!
\left\{
\mathcal{E}_{e|I\gamma},
\mathcal{E}_{I|e\gamma},
\mathcal{E}_{\gamma|eI}
\right\}.
\label{eq:MPE}
\end{equation}
where $\mathcal{E}$ denotes either $S$ or $E_N$. A pure state is genuinely multiparticle entangled precisely when it is entangled across every bipartition; consequently $\mathcal{E}_{\mathrm{GME}}>0$ provides a direct indicator of this regime \cite{huber_structure_2013,cobucci_detecting_2024}. 

The von Neumann entropies in Fig.~\ref{fig:1}(c) expose the resulting transfer particularly clearly. During photoionization, $S_{e|I\gamma}$ and $S_{I|e\gamma}$ approach unity while $S_{\gamma|eI}$ remains close to zero, corresponding to an approximately maximally entangled electron--ion state with the photon still factorized as a vacuum state, $\ket{0}$. During spontaneous decay, $S_{e|I\gamma}$ remains equal to unity, as required because $U_{I\gamma}$ is local with respect to this bipartition. At the same time, $S_{I|e\gamma}$ decreases and $S_{\gamma|eI}$ increases. The two cross near $S\simeq0.8$, but their values should not be interpreted as additive fractions of a conserved entanglement: different bipartitions contain overlapping information about the same correlations. In the intermediate region all three entropies are nonzero, giving the shaded multiparticle-entangled region. Once decay is complete, the ion factorizes in state $\ket{a}$ and $S_{I|e\gamma}\rightarrow0$, whereas the electron and photon retain approximately one bit of entanglement.

Figure~\ref{fig:1}(d) shows the same one-versus-rest partitions using logarithmic negativity, together with the three pairwise entanglements. While the logarithmic negativities reproduce the qualitative entanglement transfer found from the von Neumann entropy, we additionally find that $E_N(e|I\gamma)$ increases slightly above unity near the end of the pulse and remains above unity during the subsequent fluorescence. This increase arises from weak high-dimensional correlations associated with the photon-energy continuum, beyond an effective qubit description. 
This raises the question of why the different measures of entanglement show different sensitivity to the continuum part of the state.  
For the pure 
system 
the von Neumann entropy is \cite{haroche_exploring_2006}
\begin{equation}
S_{A|B}=-\sum_k\lambda_k\log_2\lambda_k.
\end{equation}
where $\lambda_k$ are eigenvalues of $\rho_A$ (or $\rho_B$), also known as Schmidt probabilities. 
A similar expression for the logarithmic negativity is \cite{vidal_computable_2002} \begin{equation}\label{Eq:EN_pure}
    E_N(A|B)=2\log_2\sum_k\sqrt{\lambda_k}.    
\end{equation}

Adding a weakly populated Schmidt mode to the $N$ dominant modes: $\lambda_0\ll\lambda_1,\lambda_2,...,\lambda_N$, leads to an additional entanglement that scales as $\Delta S_\text{vN}\propto-\lambda_0\log_2\lambda_0$ and $\Delta E_N\propto \sqrt{\lambda_0}$, respectively. 
Thus, it is clear that $\Delta S_\text{vN}\ll \Delta E_N$, 
which explains why the logarithmic negativity is more sensitive to weakly populated Schmidt modes. In our case, such additional weak modes are generated during the laser interaction due to Mollow-triplet-like dynamics \cite{stenquist_entanglement_2025,stenquist_mollow-like_2024}, leading to a sizeable increase in logarithmic negativity. The von Neumann entropy also includes such effects, yielding a value that is slightly larger than one, but the effect is too weak to be seen in Fig.~\ref{fig:1}(c).  
After the laser interaction, fluorescence acts locally with respect to the $e|I\gamma$ bipartition and consequently redistributes, rather than removes, the entanglement. In exact calculations this implies that the entanglement should be constant. It should be noted that the slight decrease of $E_N(e|I\gamma)$ in Fig.~\ref{fig:1}(d) is due to irreversible decay to photon number two, an effect that is beyond the scope of the present work. 

The pairwise negativities reveal how the entanglement is distributed between the individual particles. Initially, the pairwise entanglement is almost entirely between the electron and ion. We note that its maximal value cannot exceed one because the ion is a qubit in our model. During fluorescence, $E_N(e|I)$ decreases while $E_N(e|\gamma)$ increases, directly demonstrating transfer from the electron--ion to the electron--photon pair. At the same time, a transient $E_N(I|\gamma)>0$ develops, showing that the emitted photon also becomes directly entangled with the ion during the transfer process. Once spontaneous decay is complete, both $E_N(e|I)$ and $E_N(I|\gamma)$ vanish, while $E_N(e|\gamma)$ approaches the surviving one-versus-rest negativities: $E_N(e|\gamma)\simeq E_N(e|I\gamma)\simeq E_N(\gamma|eI)$. 
The transfer therefore proceeds through a genuinely multiparticle-entangled intermediate state before reducing to an entangled electron--photon pair, $\ket{\Psi}\simeq\ket{a}_I\otimes\ket{\Phi}_{e\gamma}$.


In conclusion, these results reveal the structure underlying fluorescence-mediated entanglement transfer in photoionization. Conditioning only on ionization, we find that the initially entangled electron--ion pair evolves through a genuinely multiparticle-entangled state before approaching an entangled electron--photon pair. The one-versus-rest bipartitions show how spontaneous emission redistributes entanglement within the ion--photon subsystem while preserving the electron entanglement with the combined system, whereas the pairwise logarithmic negativities resolve the transfer from electron--ion to electron--photon entanglement and reveal a transient direct entanglement between the ion and photon. Logarithmic negativity additionally exposes weak high-dimensional correlations carried by the photon-energy continuum that are suppressed in the von Neumann entropy. The resulting dynamics therefore cannot be understood as a simple transfer of a single bipartite resource, but instead proceed through a transient three-particle state in which entanglement is shared among all constituents, with additional weak high-dimensional correlations carried by the photon-energy continuum.

\section*{Acknowledgement}
This work was supported by the Knut and Alice Wallenberg Foundation: 2024.0212; and the Swedish Research Council: 2024-04247.

\bibliography{Refs}

\end{document}